\documentclass[conference]{new-aiaa}
\usepackage[utf8]{inputenc}
\usepackage{subdepth}
\usepackage{graphicx}
\usepackage{verbatim}
\usepackage{amsmath}
\usepackage[version=4]{mhchem}
\usepackage{siunitx}
\usepackage{longtable,tabularx}
\usepackage{multicol}
\usepackage{subcaption} 
\usepackage{algorithm}
\usepackage{algpseudocode}
\usepackage{todonotes}

\title{Extension of graph-accelerated non-intrusive polynomial chaos to high-dimensional uncertainty quantification through \\ the active subspace method}

\author{Bingran Wang\footnote{Corresponding author. E-mail: b6wang@ucsd.edu}, Nicholas C. Orndorff, Mark Sperry, and John T. Hwang}
\affil{University of California, San Diego, La Jolla, CA, 92093}

\begin{document}

\maketitle
 \begin{abstract}
The recently introduced graph-accelerated non-intrusive polynomial chaos (NIPC) method has shown effectiveness in solving a broad range of uncertainty quantification (UQ) problems with multidisciplinary systems. 
It uses integration-based NIPC to solve the UQ problem and generates the quadrature rule in a desired tensor structure, so that the model evaluations can be efficiently accelerated through the computational graph transformation method, Accelerated Model evaluations on Tensor grids using Computational graph transformations (AMTC).
This method is efficient when the model's computational graph possesses a certain type of sparsity which is commonly the case in multidisciplinary problems.
 However, it faces limitations in high-dimensional cases due to the curse of dimensionality.
To broaden its applicability in high-dimensional UQ problems, we propose AS-AMTC, which integrates the AMTC approach with the active subspace (AS) method, a widely-used dimension reduction technique. In developing this new method, we have also developed AS-NIPC, linking integration-based NIPC with the AS method for solving high-dimensional UQ problems. 
AS-AMTC incorporates rigorous approaches to generate orthogonal polynomial basis functions for lower-dimensional active variables and efficient quadrature rules to estimate their coefficients.
The AS-AMTC method extends AS-NIPC by generating a quadrature rule with a desired tensor structure.
This allows the AMTC method to exploit the computational graph sparsity, leading to efficient model evaluations.
In an 81-dimensional UQ problem derived from an air-taxi trajectory optimization scenario, AS-NIPC demonstrates a 30\% decrease in relative error compared to the existing methods, while AS-AMTC achieves an 80\% reduction.
\end{abstract}
\section{Introduction}
Uncertainties are ubiquitously present in many scientific applications and can significantly impact a system’s behavior or performance.
Uncertainty quantification (UQ) aims
to quantify and analyze the impact of uncertain inputs on a system’s outputs or quantities of interest (QoIs).
By providing reliable estimates of the uncertainties, UQ can support decision-making and risk assessment processes in engineering system design.
The applications of UQ span across various engineering disciplines, including structure analysis~\cite{wan2014analytical, hu2018uncertainty}, system control~\cite{michelmore2020uncertainty}, machine learning~\cite{abdar2021review}, and aircraft design~\cite{ng2016monte,wang2024graph1}.

Input uncertainties can arise from various sources and are generally categorized as aleatoric or epistemic. Aleatoric uncertainties, also known as irreducible uncertainties, are inherent to the system and may include variations in operating conditions, mission requirements, and model parameters, among others. These uncertainties can often be defined within a probabilistic framework. Epistemic uncertainties, on the other hand, are reducible uncertainties that arise from a lack of knowledge. Such uncertainties may be caused by approximations used in the computational model or numerical solution methods and are often difficult to define within a probabilistic framework. 
This paper focuses on UQ problems within a probabilistic formalism, where the uncertain inputs are continuous random variables characterized by known probability density distributions. The objective is to calculate key statistical parameters like the mean and variance of the QoI, or more intricate risk measures such as the probability of failure and the conditional value at risk.

UQ has been a major research area for many years and non-intrusive approaches have gained popularity due to their ease of implementation. 
The most common non-intrusive methods include the method of moments, polynomial chaos, kriging, and Monte Carlo.
These methods have their strengths and weaknesses, and the choice of method depends on the problem at hand.
The method of moments, employs a Taylor series approximation to analytically determine the statistical moments of the QoI using function evaluations and derivative information at one or a few predetermined points~\cite{wooldridge2001applications}. While efficient for estimating statistical moments, this method struggles with more complex risk measures and may encounter accuracy issues when dealing with large input variances.
The Monte Carlo approach involves randomly sampling the uncertain inputs and then calculating risk measures based on the resulting output data. Due to the rule of large numbers, the convergence of Monte Carlo is independent of the number of uncertain inputs, which makes it the desirable method to use for solving high-dimensional problems. 
Recent efforts to enhance its efficiency have focused on employing multi-fidelity techniques~\cite{peherstorfer2016optimal,peherstorfer2018survey} and importance sampling methods~\cite{tabandeh2022review}.
Nevertheless, for low-dimensional problems, the Monte Carlo method may necessitate a significantly larger number of model evaluations to achieve comparable accuracy to other UQ methods.
Kriging, also referred to as Gaussian process regression, creates a response surface function by leveraging input-output data. This trained kriging model serves as a surrogate for the original function, enabling numerous model evaluations for conducting reliability analyses~\cite{kaymaz2005application,hu2016single} or optimization under uncertainty~\cite{rumpfkeil2013optimizations}.
The polynomial chaos method represents the QoI as orthogonal polynomials that are derived from the distributions of the uncertain inputs. This method exploits the smoothness in the random space and enables a fast convergence rate with sampling or integration techniques.
Common polynomial chaos-based methods include non-intrusive polynomial chaos~\cite{hosder2006non,jones2013nonlinear,keshavarzzadeh2017topology} and stochastic collocation~\cite{xiu2005high, babuvska2007stochastic}.
While kriging and polynomial chaos are favored for low or medium-dimensional problems, they encounter challenges with high-dimensional UQ due to the curse of dimensionality. Their efficiency diminishes as the problem's dimension increases, making them less suitable to be directly applied in such scenarios.

Recently, Wang et al. introduced the graph-accelerated NIPC method for tackling UQ problems that involve multidisciplinary systems~\cite{wang2024accelerating, wang2024graph}. This approach combines a novel computational graph transformation technique known as \textit{Accelerated Model evaluations on Tensor grids using Computational graph transformations} (AMTC)~\cite{wang2024accelerating,wang2022efficient} with the integration-based NIPC method. 
The AMTC method accelerates tensor-grid evaluations by modifying the computational graph of the model, thereby eliminating redundant evaluations at the operation level caused by the tensor structure of the inputs. This method has been well connected to the integration-based NIPC to generate the quadrature rules with desired tensor structures to exploit sparsities in the computational models. 
While this approach has demonstrated superiority over existing UQ methods for specific low-dimensional UQ problems involving multidisciplinary systems, it suffers from the curse of dimensionality in many high-dimensional UQ problems.

When addressing high-dimensional UQ problems, employing dimension-reduction techniques can sometimes significantly reduce computational costs.
Notable methods include sensitivity analysis \cite{morio2011global}, principal component analysis (PCA)~\cite{abdi2010principal}, partial least squares (PLS)~\cite{chun2010sparse}, and active subspace (AS)~\cite{constantine2014active}.
Sensitivity analysis is a commonly used method that determines the impact of input parameters on output uncertainty. Local sensitivity analysis measures the perturbation of each input at a nominal value and its effect on output, making it inexpensive to implement. However, this method may lack accuracy in many cases. 
In contrast, global sensitivity analysis \cite{sobol2001global} computes Sobol indices, which measure output variations over the full range of inputs, providing more accurate results but at a higher computational cost.
The inputs corresponding to the highest sensitivities are then selected to reduce the dimension of the problems.
PCA conducts eigenvalue decomposition on the covariance matrix of uncertain inputs, revealing the highest-variance directions in the input space. 
Yet, PCE is not well-suited for independent uncertain inputs.
PLS utilizes regression to identify covariant directions between inputs and outputs. However, it can suffer from overfitting and does not leverage gradient information.
On the other hand, AS uses gradient information to identify an active subspace where most of the first-order changes in the output exist. AS strikes a balance between accuracy and computational cost, making it a powerful tool for high-dimensional UQ problems, while also providing an intuitive way to visualize and interpret the results.

This paper aims to extend the capabilities of the graph-accelerated NIPC methods to solve high-dimensional UQ problems by leveraging the active subspace method. While the graph-accelerated NIPC method has proven effective in solving low-dimensional UQ problems with sparsity in computational graphs, there is currently no existing framework that can directly connect this method to the AS method for solving high-dimensional UQ problems.
In previous works, Glaws and Constantines connected the AS method to the Gaussian quadrature rules to solve integration problems~\cite{glaws2019gaussian}, but this method is limited to one-dimensional active subspace. 
Additionally, in \cite{he2023novel}, He et al., linked the AS method to regression-based NIPC methods, but this approach cannot be used to connect the active subspace method to the graph-accelerated NIPC method, which relies on an integration-based approach to solve the polynomial chaos expansion (PCE) coefficients. 
To address this gap, this paper first proposes a general framework called AS-NIPC, which generates PCE basis functions for the active subspace's uncertain inputs and an efficient quadrature rule in the original uncertain input space to connect the active subspace method to the integration-based NIPC method. 
This framework can be applied to any high-dimensional UQ problem without using the AMTC method. 
Furthermore, an extension of this framework, AS-AMTC, is presented to use with the AMTC method. This extension involves identifying the sparse uncertain inputs (uncertain inputs that only affect a small amount of the computational cost), applying the AS method to non-sparse uncertain inputs, and generating a desired tensor-structured quadrature rule to take advantage of the computational graph sparsity when the AMTC method is applicable.

The proposed methods have been tested on two UQ problems which include a 7-dimensional analytical piston model and an 81-dimensional air-taxi trajectory simulation model.
The results show that the proposed AS-NIPC is more effective than the existing methods for both of the problems while AS-AMTC can further improve its efficiency on the second problem which involves a multidisciplinary system.

This paper is organized as follows. 
Section \ref{Sec: Background} gives some background for graph-accelerated NIPC and AS methods. 
Section \ref{Sec: Methodology} presents the details of the AS-NIPC and AS-AMTC methods. 
Section \ref{Sec: Numerical Results} shows the numerical results of the test problems.
Section \ref{Sec: Conclusion} summarizes the work and offers concluding thoughts.

\section{Background}
\label{Sec: Background}

\subsection{Non-intrusive polynomial chaos}
The generalized polynomial chaos (gPC)~\cite{xiu2002wiener}, proposed by Xiu and Karniadakis, is a widely used method to address UQ problems with continuous random inputs. The key idea of gPC is to approximate the stochastic output of a model as a linear combination of orthogonal polynomial basis functions that are chosen according to the distributions of the uncertain inputs.

We address an uncertainty quantification (UQ) problem, involving a function represented as:
\begin{equation}
f = \mathcal{F}(u),
\end{equation}
where $\mathcal{F}:\mathbb{R}^{d} \to \mathbb{R}$ denotes a deterministic model evaluation function, $u \in \mathbb{R}^d$ represents the input vector, and $f \in \mathbb{R}$ represents a scalar output. 
The input uncertainties are denoted as the stochastic vector $U := [U_1, \ldots, U_d]$ and the UQ problem aims to compute the statistical moments or risk measures of the stochastic output $f(U)$. The uncertain inputs are assumed to be mutually independent and are associated with the joint probability density distribution $\rho(u)$ with support $\Gamma$.

In gPC, the stochastic output can be represented an infinite series of orthogonal polynomials of the uncertain inputs:
\begin{equation}
\label{eqn: pce}
    f(U) = \sum_{i = 0}^{\infty} \alpha_i \Phi_i(U),
\end{equation}
where $ \Phi_i(U)$ and $\alpha_i$ are polynomial chaos expansion (PCE) basis functions and their corresponding coefficients, respectively.
In practice, \eqref{eqn: pce} is truncated to a certain order, leading to an approximation function, written as
\begin{equation}
\label{Eqn: pce estimate}
    f(U) \approx \sum_{i = 0}^{q} \alpha_i \Phi_i(U).    
\end{equation}
When truncating the series to include only polynomials up to order $p$, the total number of PCE terms, $q + 1$, is
\begin{equation}
    q + 1 = \frac{(d + p)!}{d
    ! p!}.
\end{equation}
These PCE basis functions are chosen to satisfy the orthogonality 
property:
\begin{equation}
\label{Eqn: orthogonal property}
    \left< \Phi_i(U), \Phi_j(U) \right> = \delta_{ij},
\end{equation}
where $\delta_{ij}$ is Kronecker delta and the inner product is defined as:
\begin{equation}
    \left< \Phi_i(U), \Phi_j(U) \right> = \int_{\Gamma}  \Phi_i(u) \Phi_j(u)\rho(u) du.
\end{equation}

Selecting the right PCE basis functions is crucial for ensuring the rapid convergence of any gPC-based methods, a task that can be complex in various scenarios. In one-dimensional cases, univariate orthogonal polynomials have been derived for common types of continuous random variables, as detailed in Table~\ref{tab: orthog poly}. For multi-dimensional problems with mutually independent inputs, the PCE basis functions can be constructed as a tensor product of the univariate orthogonal polynomials corresponding to each uncertain input.
However, this process does not work directly if the uncertain inputs are not independent or do not follow standard distributions. If the uncertain inputs are multivariate Gaussian random variables, the PCE basis functions can be derived analytically following methods outlined in \cite{rahman2017wiener}.
For uncertain inputs following arbitrary distributions, PCE basis functions can be numerically generated using a whitening transformation matrix approach following \cite{lee2020practical}.

There are mainly two approaches to solve the PCE coefficients in \eqref{Eqn: pce estimate}: by integration or regression. These two approaches show similar performance on low-dimensional UQ problems~\cite{eldred2009comparison}, but their research focuses are drastically different.
The integration approach makes use of the orthogonality property of PCE basis functions in \eqref{Eqn: orthogonal property} and projects the QoI onto each basis function, which results into an integration problem for each coefficient:
\begin{equation}
\begin{aligned}
\label{eqn: pce_integration}
    \alpha_i & = \frac{\left< f(U), \Phi_i \right> }{\left< \Phi_i^2 \right>} \\
    & =\int_{\Gamma} f(u) \Phi_i(u) \rho(u) du\\
    & = \int_{\Gamma_1}\ldots \int_{\Gamma_d}  f(u_1, \ldots, u_d) \Phi_i(u_1,\ldots, u_d)\rho(u_1,\ldots, u_d)du_1 \ldots du_d. \\
\end{aligned}
\end{equation}
In this approach, the number of the PCE basis functions we use does not affect the model evaluation cost as we use the same quadrature rule to estimate all of the PCE coefficients.
The main difficulty here is to derive an efficient quadrature rule to solve the high-dimensional integration problem in \eqref{eqn: pce_integration}. 
The most straightforward quadrature rule to use is the Gauss quadrature rule~\cite{golub1969calculation}. 
In 1-dimensional case, the Gauss quadrature rule is optimal to use as the quadrature rule with $k$ nodes can exactly integrate polynomials up to ($2k-1$)th order. However, in the multi-dimensional case, the quadrature rule is formed as a tensor product of the 1D quadrature rule and the number of quadrature points increases exponentially with the dimension as $k^d$. More efficient quadrature rules to use for high-dimensional problems include Smolyak sparse-grid quadrature rule~\cite{smolyak1963quadrature} which drops higher order terms in Gauss quadrature rule while maintaining minimal loss and the designed quadrature rule~\cite{keshavarzzadeh2018numerical}, which numerically generate the quadrature rule that satisfies the moment matching equations through optimization.

On the other hand, the regression approach computes the PCE coefficients by solving a least-squares linear system:
\begin{equation}
    \begin{bmatrix}
    \Phi_1(u^{(1)}) &  \ldots & \Phi_q(u^{(1)}) \\
    \vdots & & \vdots \\
    \Phi_1(u^{(n)}) &  \ldots & \Phi_q(u^{(n)}) \\
    \end{bmatrix}
    \begin{bmatrix}
    \alpha_1 \\
    \vdots \\
    \alpha_q 
    \end{bmatrix} =
    \begin{bmatrix}
    f(u^{(1)}) \\
    \vdots \\
    f(u^{(n)})
    \end{bmatrix},
\end{equation}
where $(u^{(1)},\ldots, u^{(n)})$ denotes a set of $n$ sample points generated from the distributions of the uncertain inputs. Typically, the number of samples $n$ is chosen to be 2-3 times the number of coefficients.
This means the number of PCE basis functions used in NIPC would directly affect the required number of model evaluations and the accuracy of the NIPC method. 
Thus the main avenues for improving its efficiency are through sparse PCE basis selection~\cite{blatman2008sparse,tipireddy2014basis} and optimal sampling~\cite{hampton2015compressive}.

\begin{table}[]
\caption{Orthogonal polynomials for common types of continuous random variables}
\centering
\begin{tabular}{c c c c} 
 \hline
 Distribution & Orthogonal polynomials & Support range \\ 
 \hline\hline
 Normal & Hermite  & $(-\infty, \infty)$ \\
 \hline
 Uniform & Legendre  & $[-1, 1]$  \\
 \hline
 Exponential & Laguerre  & $[0, \infty)$ \\
 \hline
 Beta & Jacobi  & $(-1, 1)$ \\
 \hline
 Gamma & Generalized Laguerre  & $[0, \infty)$ \\
 \hline
\end{tabular}
\label{tab: orthog poly} 
\end{table}

\subsection{Computational graph transformations to accelerate non-intrusive polynomial chaos}
\begin{figure}%
    \centering
    \subfloat[\centering Computational graph without using AMTC]{{\includegraphics[width=5cm]{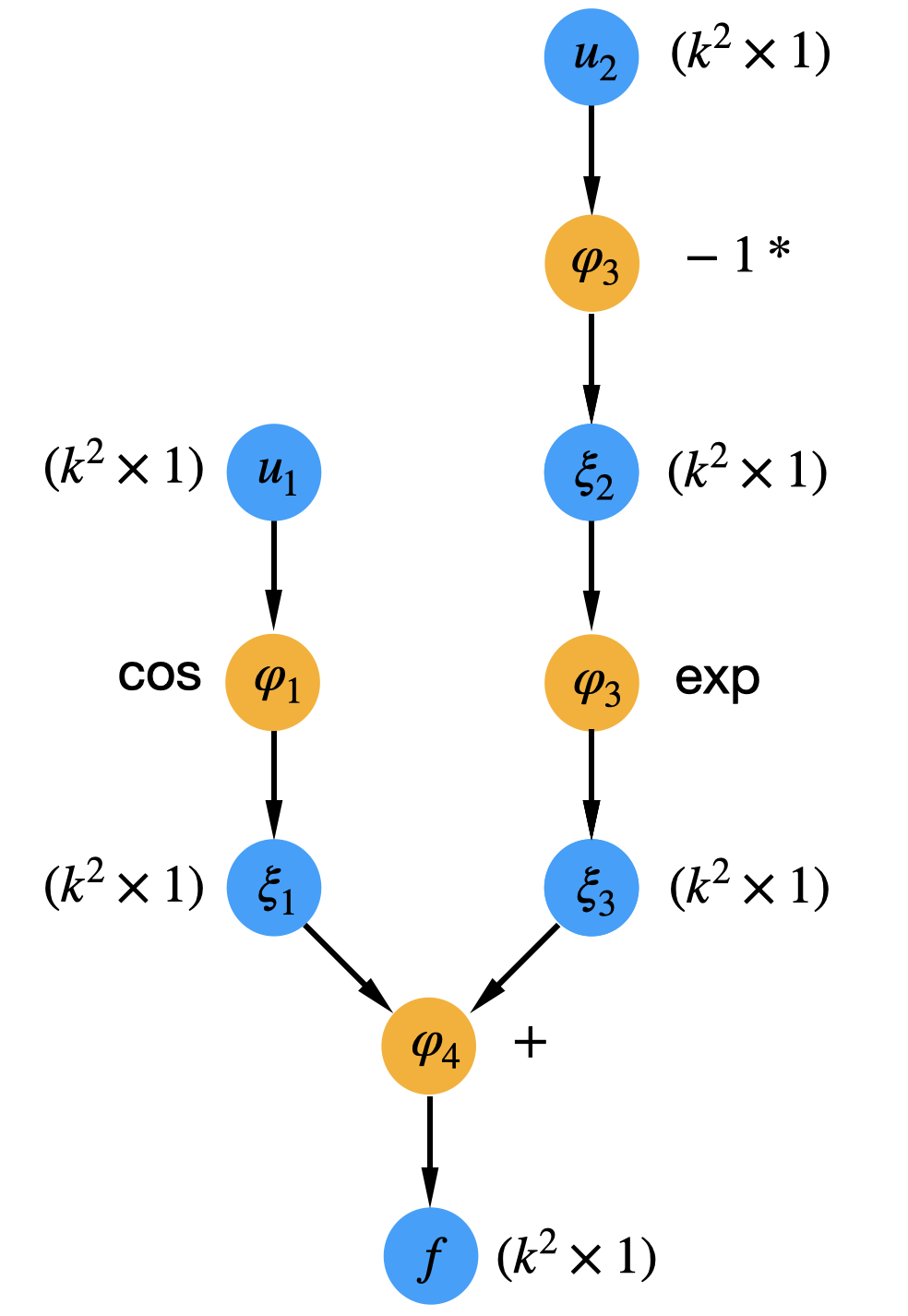} }}%
    \qquad
    \subfloat[\centering Computational graph using AMTC]{{\includegraphics[width=5cm]{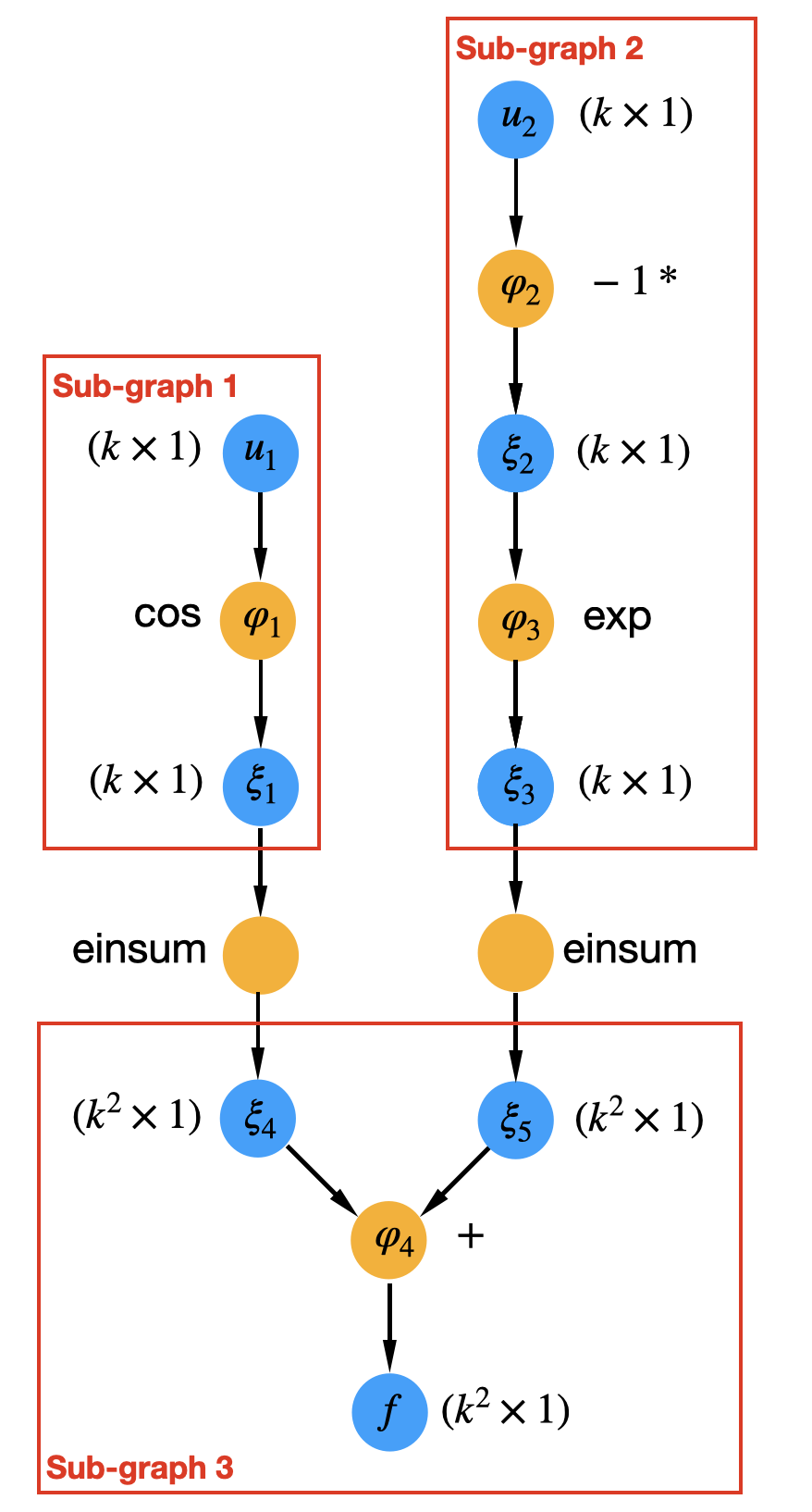} }}%
    \caption{Computational graphs with data size for full-grid input points evaluation on $f = cos(u_1) + exp(-u_2)$}%
    \label{fig:graph comparison}%
\end{figure}

Recently, Wang et al. introduced the \textit{Accelerated Model evaluations on Tensor grids using Computational graph transformations} (AMTC)~\cite{wang2024accelerating} method to reduce the model evaluations cost on tensor structured inputs via a computational graph transformation approach.
The AMTC method takes advantage of the fact that when we view a computational model as a computational graph with elementary operations, each operation's output only requires distinct evaluations on the distinct nodes in the space of its dependent inputs. 
When evaluating a model on tensor structured input nodes, often many operations do not depend on all of the uncertain inputs and the traditional framework to run the entire model in a for-loop approach creates many repeated evaluations on the operation level.
To address this, the AMTC method partitions the computational graph into sub-graphs based on the dependency information of the operations.
 Operations within each sub-graph share the same input space and are evaluated only on the distinct nodes within that space. 
Additionally, the Einstein summation operations are inserted connecting the sub-graphs to ensure correct data flow between the sub-graphs.
For example, suppose we want to evaluate a simple function
\begin{equation}
\label{eqn: simple func}
    f = cos(u_1) + exp(-u_2)
\end{equation}
on full-grid input points with $k$ points in each dimension. The function can be decomposed into a sequence of elementary operations as
\begin{equation}
\label{eqn: computational process}
    \begin{aligned}
    & \xi_1 =  \varphi_1 (u_1) = cos (u_1); \\
    & \xi_2 = \varphi_2 (u_2) =  - u_2; \\
    & \xi_3 =  \varphi_3 (\xi_2) =  exp (\xi_2); \\
    & f = \varphi_4 (\xi_2, \xi_3) =  \xi_2 + \xi_3, \\
    \end{aligned}
\end{equation}
where $\varphi_i$ represents the elementary operations in the function. 
The input points are given as
\begin{equation}
\boldsymbol{u} = \begin{Bmatrix}
(u_1^{(1)}, u_2^{(1)}) & \ldots & (u_1^{(k)}, u_2^{(1)}) \\
\vdots & \ddots & \vdots \\
(u_1^{(1)}, u_2^{(k)}) & \ldots & (u_1^{(k)}, u_2^{(k)}) \\
\end{Bmatrix},
\end{equation}
with $k^2$ input points. The computational graphs for this function with and without the AMTC method are shown in Fig.~\ref{fig:graph comparison}.
When not using the AMTC method, all of the operations in the computational graph are evaluated $k^2$ times.
However, when using the AMTC method, the operations that are only dependent on $u_1$ or $u_2$ are evaluated $k$ times as their outputs only have $k$ distinct values. 

\begin{figure}
\centering
  \includegraphics[width=14cm]{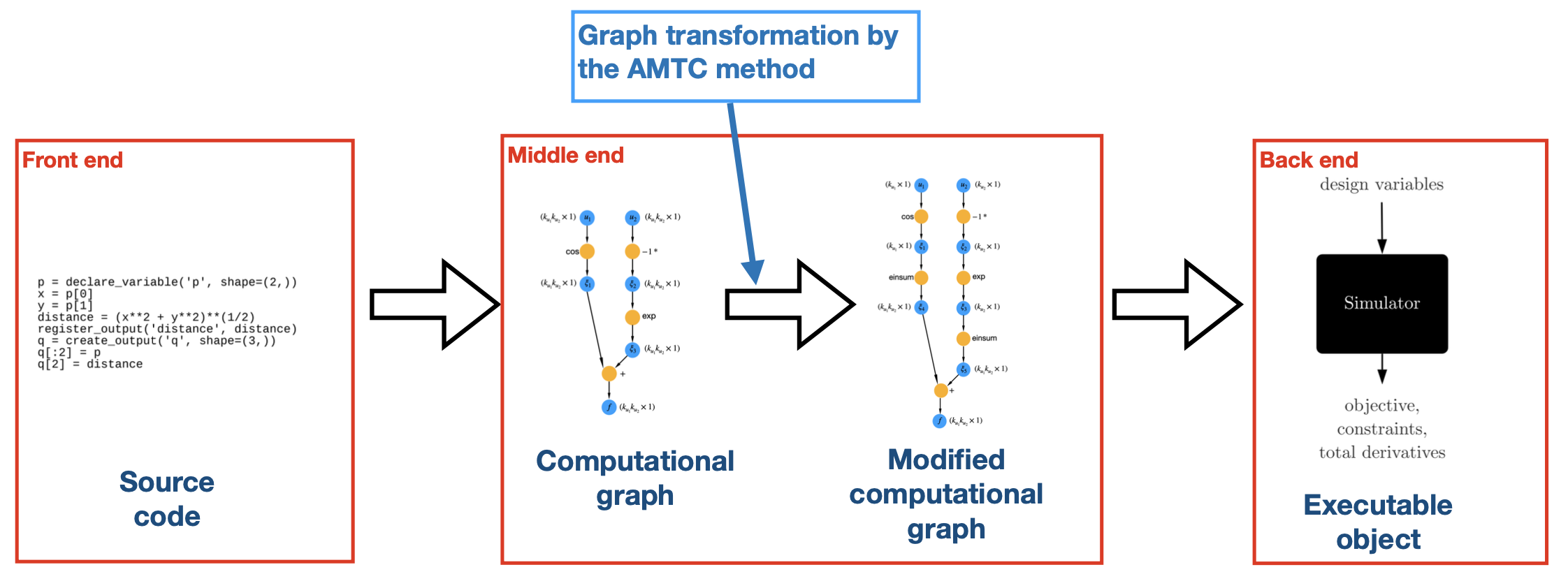}
\caption{Demonstration for the AMTC implementation in CSDL}
\label{fig:AMTC_CSDL}
\end{figure}

This method has been implemented in the middle end of the compiler for a recently developed modeling language, \textit{Computational System Design Language} (CSDL)~\cite{gandarillas2022novel}. CSDL is a domain-specific language designed for large-scale multidisciplinary design, analysis, and optimization and generates the computational graph of the model based on the source code. When evaluating the model on tensor structured inputs, the AMTC method is used to generate a modified computational graph so that the repeated evaluations on the operation level are eliminated. A demonstration for the AMTC implementation in CSDL is shown in Fig.~\ref{fig:AMTC_CSDL}.

The AMTC method has been well-connected to the integration-based NIPC approach, which uses full-grid quadrature rules to solve low-dimensional UQ problems. 
This is known as the graph-accelerated NIPC method, which has demonstrated effectiveness in handling various low-dimensional UQ problems with multidisciplinary systems.
To expand the applications of graph-accelerated NIPC method into higher dimensional UQ problems, in~\cite{wang2024graph}, Wang et al. proposed a framework to generate a partially tensor-structured quadrature rule to use within this method.
This approach involves constructing a quadrature rule in a desired tensor structure by analyzing the computational graph and generating the quadrature rule using the designed quadrature approach.
By leveraging the AMTC method, the corresponding model evaluations cost can be minimized, effectively utilizing the sparsity inherent in the computational graph.

\subsection{Active subspace}
The active subspace (AS) method, proposed by Constantine et. al. in~\cite{constantine2014active}, is a dimension reduction method that has recently gained a lot of popularity. 
The basic idea behind AS is to exploit the fact that some directions in the input parameter space have a greater impact on the output uncertainty than others. By identifying these important directions, we can project the high-dimensional input space onto a lower-dimensional subspace that captures the most important first-order input-output relationships. To achieve that, the AS method computes the 
mean squared gradients of the objective function with respect to the uncertain inputs as
\begin{equation}
    C = \mathbb{E} [(\nabla_u f)(\nabla_u f)^T] =  \int_{\Gamma} (\nabla_u f)(\nabla_u f)^T\rho(u) du\\,
\end{equation}
where $\nabla_u f = [\frac{\partial f}{\partial u_1}, \ldots, \frac{\partial f}{\partial u_d}]^T$. The $C$ matrix is symmetric and positive semi-definite and thus has non-negative eigenvalues and orthogonal eigenvectors. Following an eigenvalue decomposition, the $C$ matrix can be expressed as
\begin{equation}
    C = W \Lambda W^T,\quad \Lambda = \text{diag}(\lambda_1,\ldots, \lambda_d ).
\end{equation}
The eigenvectors $W$ define a rotation of the original uncertain input space and are sorted by the magnitude of their corresponding eigenvalues in decreasing order, while the magnitude of the eigenvalues represents the function's average variation in each direction. The AS method identifies the active subspace $\Tilde{u}$ by partitioning the eigenvalues and eigenvectors as
\begin{equation}
\Lambda = \begin{bmatrix}
\Lambda_1 & \\
 & \Lambda_2
\end{bmatrix}, \quad  W =  \begin{bmatrix}
W_1 & W_2\\
\end{bmatrix},       
\end{equation}
where  $\Lambda_1= \text{diag} (\lambda_1,\ldots, \lambda_m)$ and $W_1 \in \mathbb{R}^{m\times d}$ with $m < d$. This results in two sets of rotated coordinates defined as
\begin{equation}
    \Tilde{u} = W_1^Tu, \quad \bar{u} = W_2^Tu,
\end{equation}
such that the function is significantly more variant in the space of $\Tilde{u}$ than $\bar{u}$. 
Once the active subspace is identified, the UQ problem can be solved in this lower-dimensional space, which greatly reduces the computational cost of the analysis.
\section{Methodology}
\label{Sec: Methodology}

In this paper, we first propose a novel framework to connect the integration-based NIPC method to the active subspace approach to solve high-dimensional UQ problems. This involves generating orthogonal PCE basis functions of the active subspace variables using the whitening matrix and generating an efficient quadrature rule in the original input space using the designed quadrature method to estimate the PCE coefficients.  We then propose an extension of this framework to combine it with the AMTC method to generate the quadrature rule that has a desired tensor structure to take advantage of the sparsity of the computational graph.
\subsection{AS-based NIPC}
Consider a high-dimensional UQ problem involving a function,
\begin{equation}
    f = \mathcal{F}(u) \quad u \in \mathbb{R}^d,
\end{equation}
with the independent uncertain inputs $U = (U_1,\ldots, U_d)\in \mathbb{R}^d$.
We first apply the active subspace (AS) method to find a desired rotation of the input coordinate such that the function is only significantly variant with respect to a small number of inputs.
This involves first computing the mean squared gradients of the objective function with respect to the uncertain inputs as
\begin{equation}
\label{eqn: 1}
    C = \mathbb{E} [(\nabla_u f)(\nabla_u f)^T],
\end{equation}
followed by an eigenvalue decomposition as
\begin{equation}
\label{eqn: 2}
    C = W \Lambda W^T, \quad \Lambda = \text{diag}(\lambda_1,\ldots, \lambda_d ).
\end{equation}
The eigenvalues and eigenvectors are then partitioned as
\begin{equation}
\label{eqn: 3}
\Lambda = \begin{bmatrix}
\Lambda_1 & \\
 & \Lambda_2
\end{bmatrix}, \quad  W =  \begin{bmatrix}
W_1 & W_2\\
\end{bmatrix}   
\end{equation}
with $\Lambda_1 = \text{diag}(\lambda_1,\ldots, \lambda_m)$. This separates the rotated coordinates into two sets as
\begin{equation}
\label{eqn: 4}
    \Tilde{u} = W_1^Tu, \quad \bar{u} = W_2^Tu,
\end{equation}
and the function varies significantly more with respect to $\Tilde{u}$ than $\bar{u}$.
We call $\Tilde{u}$ the \textit{active variables} and $\Bar{u}$ the \textit{inactive variables}.
Now the function can be written as 
\begin{equation}
    f (\Tilde{u},\bar{u} ) =  \mathcal{F}(W_1 u + W_2u) .
\end{equation}
By using the gPC theory and truncating all of the terms of $\bar{u}$, we approximate the model output as
\begin{equation}
\label{eqn: as_nipc}
    f(\Tilde{U}) \approx \sum_{i = 0}^{q} \alpha_i \Phi_i(\Tilde{U}).
\end{equation}
Since $\Tilde{u}$ can be significantly lower dimensional than the original uncertain input space, $u$, solving the PCE coefficients in equation~\eqref{eqn: as_nipc} can be significantly less expensive than directly applying NIPC. 
However, applying the integration-based NIPC method here poses two main challenges: 
\begin{enumerate}
    \item The probability density of the distribution of the uncertain inputs in the active space is unknown which makes it difficult to generate the PCE terms in the active subspace.
    \item An efficient quadrature rule in the original input space needs to be derived to compute the PCE coefficients that correspond to the PCE terms of the active subspace uncertain inputs.
\end{enumerate}
\subsubsection{Generating PCE basis functions}
We solve the first challenge by computing the whitening matrices that represent the coefficients of the PCE basis functions. 
We follow the method in~\cite{lee2020practical} and first represent the PCE basis functions of $\Tilde{u}$ as
\begin{equation}
\label{eqn: 5}
     \Phi(\Tilde{u}) =
    \begin{bmatrix}
        \Phi_0(\Tilde{u}), \ldots,  \Phi_q(\Tilde{u})
    \end{bmatrix}^T = M P_k(\Tilde{u}),
\end{equation}
where  $P_k(\Tilde{u})$ is the monomial basis vector in ${\Tilde{u}}$ up to degree $k$, and $M \in \mathbb{R}^{(k+1)\times(k+1)}$ is a lower triangular matrix storing the coefficients for the monomial basis polynomials in the univariate PCE basis functions. 
For demonstration, in a two-dimensional case, the first six Legendre polynomials (PCE basis functions for uniform uncertain inputs) are
\begin{equation}
\label{eqn: lagrendre_pce}
    \begin{aligned}
    & \Phi_0(u) =  1; \\
    & \Phi_1(u) = u_1; \\
    & \Phi_2(u) = u_2; \\
    & \Phi_3(u) = u_1u_2; \\
    & \Phi_4(u) = \frac{1}{2} (3 u_1^2-1); \\
    & \Phi_5(u) = \frac{1}{2} (3 u_2^2-1). \\
    \end{aligned}
\end{equation}
The PCE basis functions in \eqref{eqn: lagrendre_pce} can be represented in the form of the monomial basis vector as
\begin{equation}
    \underbrace{\begin{bmatrix}
        \Phi_0(u) \\ \Phi_1(u) \\ \Phi_2(u) \\ \Phi_3(u) \\ \Phi_4(u) \\  \Phi_5(u) \\
    \end{bmatrix}}_{\Phi(u)} = \underbrace{\begin{bmatrix}
1 & 0 & 0 & 0 & 0 & 0 \\
0 & 1 & 0 & 0 & 0 & 0 \\
0 & 0 & 1 & 0 & 0 & 0 \\
0 & 0 & 0 & 1 & 0 & 0 \\
-\frac{1}{2} & 0 & 0 & 0 & \frac{3}{2} & 0 \\
-\frac{1}{2} & 0 & 0 & 0 & 0 & \frac{3}{2} \\
\end{bmatrix}}_M
\underbrace{\begin{bmatrix}
1 \\ u_1 \\ u_2 \\ u_1u_2 \\ u_1^2 \\ u_2^2 \\
\end{bmatrix}}_{P_2(u)}
\end{equation}
We want to solve for the $M$ matrix so that the orthogonality property in \eqref{Eqn: orthogonal property} is satisfied in terms of $\Tilde{u}$ as
\begin{equation}
    \int_{\Gamma_{\Tilde{u}}} \Phi(\Tilde{u}) \Phi(\Tilde{u})^T\rho(\Tilde{u}) d\Tilde{u} = I, 
\end{equation}
which can be written as
\begin{equation}
    \int_{\Gamma_{\Tilde{u}}} P_k(\Tilde{u}) P_k(\Tilde{u})^T\rho(\Tilde{u})d\Tilde{u} = M^{-1}M^{-T}.
\end{equation}
We define the monomial matrix as
\begin{equation}
\label{eqn: G}
    G: = \int_{\Gamma_{\Tilde{u}}} P_k(\Tilde{u}) P_k(\Tilde{u})^T\rho(\Tilde{u})d\Tilde{u},
\end{equation}
then the $M$ matrix satisfies
\begin{equation}
    M^{-1}M^{-T} =  G,
\end{equation}
and can be solved through Cholesky factorization of $G$ as
\begin{equation}
\label{eqn: 6}
    M =  Q^{-1}, \quad G = Q Q^T.
\end{equation}
However, solving $G$ in \eqref{eqn: G} would require us to know the probability distribution of the active variables, $\rho(\Tilde{u})$. 
To avoid solving $\rho(\Tilde{u})$, we utilize the linear relationship, $\Tilde{u} =  W_1^T u$ and apply the change of variables to rewrite \eqref{eqn: G} as
\begin{equation}
\label{eqn: G_int}
\begin{aligned}
    G & = \int_{\Gamma_{\Tilde{u}}} P_k(\Tilde{u}) P_k(\Tilde{u})^T\rho(\Tilde{u})d\Tilde{u}\\
    & = \mathbb{E}[P_k(\Tilde{u}) P_k(\Tilde{u})^T] \\
    & = \mathbb{E}[P_k(W_1^T u) P_k(W_1^T u)^T] \\
    & = \int_{\Gamma_{u}} P_k( W_1^T u) P_k( W_1^T u)^T\rho(u)du.
\end{aligned}
\end{equation}
The integral in~\eqref{eqn: G_int}  does not require any model evaluation and can then be easily computed by using a Monte Carlo or numerical quadrature approach.

\subsubsection{Generating an efficient quadrature rule}
Using the integration-based NIPC approach, we approximate the PCE coefficients in \eqref{eqn: as_nipc} by solving an integration problem as
\begin{equation}
\begin{aligned}
\label{eqn: as_nipc_integration}
    \alpha_i & = \frac{\left< f(\Tilde{U}), \Phi_i(\Tilde{u}) \right> }{\left< \Phi_i(\Tilde{u})^2 \right>} \\
    & =\int_{\Gamma_{\Tilde{u}}} f(\Tilde{u}) \Phi_i(\Tilde{u}) \rho(\Tilde{u}) d\Tilde{u}.\\
\end{aligned}
\end{equation}
This results in an integration problem in $\Tilde{u}$ space, which is much lower dimensional than the original uncertain input space. However, evaluating this integral is challenging in two parts. 
Firstly, evaluating $f(\Tilde{u})$ directly is not possible as the computational model only accepts inputs in terms of $u$. Moreover, for a given point in the $\Tilde{u}$ space, there may exist an infinite number of points in $u$ that satisfy $\Tilde{u} = W_1^T u$.
Secondly, the probability density distribution of the active variables, denoted as $\rho(\Tilde{u})$, is unknown. Consequently, direct generation of quadrature points is not feasible.
The objective here is to generate an efficient quadrature rule with the nodes in the original input space.
We denote the quadrature points and the weights as $\boldsymbol{u}: = (u^{(1)},\ldots, u^{(n)})$ and $\boldsymbol{w}: = (w^{(1)},\ldots, w^{(n)})$ respectively, and the quadrature rule approximates the integral in \eqref{eqn: as_nipc_integration} as 
\begin{equation}
\label{eqn: integration}
    \int_{\Gamma_{\Tilde{u}}} f(\Tilde{u}) \Phi_i(\Tilde{u}) \rho(\Tilde{u}) d\Tilde{u} 
    \approx \sum_{i=1}^{d} w^{(i)}f(u^{(i)}) \Phi_i\left(W_1^T (u^{(i)})\right).
\end{equation}
We solve the quadrature rule following the designed quadrature method in~\cite{keshavarzzadeh2017topology}. The designed quadrature method generates the quadrature rule by solving an optimization problem to ensure they can exactly integrate polynomial functions up to a specific order. 
The core idea here is that for an integration problem 
\begin{equation}
    I(f) = \int_{\Gamma}f(u) \rho(u)du,
\end{equation}
if the nodes and weights of the quadrature rule satisfy the multi-variate moment matching equations as
\begin{equation}
\label{eqn: multi_mm}
    \sum_{i=1}^{n} \Phi_{\boldsymbol{i^\prime}} \left(u^{(i)}\right) w^{(i)}=\left\{\begin{array}{ll}
1 & \text { if } |\boldsymbol{i^\prime}|=0 \\
0 & \text { for } \alpha \in 0<|\boldsymbol{i^\prime}|<k,
\end{array}\right.
\end{equation}
where $\Phi_{\boldsymbol{i^\prime}}$ are the PCE basis functions of $u$, then the quadrature rule can exactly integrate any polynomials of $u$ up to $(2k-1)$th order.
In our case, we want to design the quadrature rule with $\boldsymbol{u}$ and $\boldsymbol{w}$ such that after the linear transformation, $\Tilde{u} = W_1^Tu$, the quadrature rule can exactly integrate polynomials of $\Tilde{u}$ to a specific order. Thus we write the moment-matching equation here as
\begin{equation}
\label{eqn: as_multi_mm}
    \sum_{i=1}^{n} \Phi_{\boldsymbol{i^\prime}} \left(W_1^Tu^{(i)}\right) w^{(i)}=\left\{\begin{array}{ll}
1 & \text { if } |\boldsymbol{i^\prime}|=0 \\
0 & \text { for }  0<|\boldsymbol{i^\prime}|<k,
\end{array}\right.
\end{equation}
where $\Phi_{\boldsymbol{i^\prime}}$ are the PCE basis functions of $\Tilde{u}$ that we generated in the previous step.
We represent the moment matching equations in \eqref{eqn: as_multi_mm} as residual equations:
\begin{equation}
    \mathcal{R} (\boldsymbol{u},\boldsymbol{w}) = 0, 
\end{equation}
where $\mathcal{R}:\mathbb{R}^{d \times n} \times \mathbb{R}^{n}  \to \mathbb{R}^n$. The quadrature rule is generated by solving an optimization problem formulated as
\begin{equation}
\label{eqn: optimization}
    \begin{aligned}
\min _{\boldsymbol{u}, \boldsymbol{w}} &\|r (\boldsymbol{u}, \boldsymbol{w})\|_{2} \\
\text { subject to } & u^{(j)} \in \Gamma, \quad j=1, \ldots, n, \\
& w^{(j)}> 0 , \quad j=1, \ldots, n,
\end{aligned}
\end{equation}
which minimizes the norm of the residual equations and constrains the nodes to be within the support range and weights to be positive. The number of nodes to use, $n$, depends on the dimension of the active subspace, $m$, and the highest degree of polynomials of PCE basis functions, $k$, and can be chosen following~\cite{wang2024graph}, as
\begin{equation}
\label{eqn: opt. num}
    n = \left\{\begin{array}{ll}
\frac{k^m}{m} & \text { for } m \leq 2 \\
0.9\frac{(2m)^{k-1}}{(k-1)!} & \text { for } m > 2.
\end{array}\right.
\end{equation}
The optimization problem in \eqref{eqn: optimization} can be solved using any common nonlinear optimizer. 
\subsubsection{Special case: Gaussian uncertain inputs}
One special case in which we can easily generate the PCE basis functions is when all of the uncertain inputs are Gaussian random variables.
The rotated coordinates $\Tilde{u}$ satisfy a linear relationship with the original coordinates $u$ as $\Tilde{u} =  W_1^T u$. 
When the uncertain inputs are mutually independent Gaussian random inputs, since the eigenvectors in $W_1$ are orthogonal, the random variables associated with the active variables are also mutually independent random variables and their distributions can be represented as
\begin{equation}
    \Tilde{U} \sim \mathcal{N}(W_1^T \mu,W_1^T\Sigma W_1),
\end{equation}
where $\mu$ and $\Sigma$ are the mean vector and covariance matrix associated with the original uncertain inputs $U$, respectively.
This means that the PCE basis functions of $\Tilde{u}$ are simply the multivariate Hermite polynomials that can be easily generated.
For the quadrature rule, we still recommend using the same designed quadrature approach, as this approach results generally require the fewest quadrature points to achieve the same level of accuracy. 
Alternatively, if the dimension of the uncertain inputs is small enough ($m \leq 3$), one may choose to use the full-grid quadrature rule in $\Tilde{u}$ and solve the quadrature points in the original input space such that $ W_1^T\boldsymbol{u} =  \boldsymbol{\Tilde{u}}$.
\subsubsection{AS-NIPC algorithm}
We refer to this AS-based NIPC method as the AS-NIPC method and present a detailed step-by-step algorithm:
\begin{enumerate}
    \item \textbf{Discover the active subspace:} Apply the AS method to discover the active variables $\Tilde{u}$, following~\eqref{eqn: 1}\eqref{eqn: 2}\eqref{eqn: 3}\eqref{eqn: 4}.
    \item \textbf{Generate the PCE basis functions:} Solve the whitening matrix, $M$, following \eqref{eqn: G_int} \eqref{eqn: 6} \eqref{eqn: 5} for a specified polynomial order $k$.
    \item \textbf{Generate the quadrature rule:} Compute the nodes $\boldsymbol{u}$ and weights $\boldsymbol{w}$ in the quadrature rule by solving the optimization problem in \eqref{eqn: optimization}, choose the number of quadrature points, $n$, using \eqref{eqn: opt. num}.
    \item \textbf{Compute the PCE coefficients:} Evaluating the model/function on the quadrature points and computing each PCE coefficient following \eqref{eqn: as_nipc_integration}\eqref{eqn: integration}.
    \item \textbf{Compute the QoIs:} Computing the desired quantity of interests of the model output following the NIPC method.
\end{enumerate}

\subsection{AS based NIPC with AMTC}

The AS-NIPC method we presented in the previous section provides a framework to combine the integration-based NIPC method with the active subspace method to solve high-dimensional UQ problems. However, this method cannot be accelerated by the AMTC method as the quadrature points do not possess a tensor structure.  
The objective here is to propose an extension of the AS-NIPC method to generate the quadrature points that possess a desired tensor structure such that the model evaluations can be significantly accelerated using the AMTC method.
We follow the partially tensor-structured quadrature rule in~\cite{wang2024graph} to generate a desired tensor-structured quadrature rule to use in the graph-accelerated NIPC method.

For high-dimensional UQ problems, there often exist some uncertain input operations that
affect a small number of operations in the computational graph. In this case, the evaluation cost of the operations that are dependent on these inputs may only take a small percentage of the total model evaluation cost.
We refer to these uncertain inputs as \textit{sparse uncertain inputs} and the other uncertain inputs as \textit{non-sparse uncertain inputs}.
The sparse uncertain inputs can be identified by computing the sparsity ratio (SR) of each uncertain input. 
SR($u_i$) is defined as the ratio of the evaluation cost of the entire model to that of only the operations that are influenced by random input $u_i$ and can be estimated based on the computational graph analysis detailed in~\cite{wang2024graph}.
We choose the sparse uncertain inputs as its sparsity ratio is $< 5\%$, and we partition the uncertain inputs as 
\begin{equation}
    u  = (u_{s}, u_{ns}), 
\end{equation}
where $u_{s}  = (u_{s_1}, \ldots, u_{s_{d_1}})\in \mathbb{R}^{d_1}$ is the set of sparse uncertain inputs, and $u_{ns} \in \mathbb{R}^{d_2}$ is the set of non-sparse uncertain inputs with $d = d_1 + d_2$. 
Since the sparse uncertain inputs only affect a small amount of model evaluation cost, we want to choose the quadrature rule that maintains a tensor structure between the quadrature points in the non-sparse uncertain input space and quadrature points in the space of each sparse uncertain input, such that the quadrature points can be written as
\begin{equation}
\boldsymbol{u} = 
\boldsymbol{u}^{n_1}_{\{ns\}}
\times
\boldsymbol{u}^{n_2}_{\{s_1\}}\times \ldots,\boldsymbol{u}^{n_2}_{\{s_{d_1}\}},
\end{equation}
where $\boldsymbol{u}^{n_1}_{\{ns\}}$ represents quadrature points in $u_{ns}$ space with $n_1$ points and $\boldsymbol{u}^{n_2}_{\{s_{d_1}\}}$ represent the quadrature points in $u_{{s_i}}$ space with $n_2$ points.
This results in a total of $n_1n_2^{d_2}$ quadrature points in the quadrature rule.
However, employing the AMTC method to transform the computational graph leads to a computational cost on the modified graph that scales roughly linearly with $n_1$. This is because most operations in the computational graph depend solely on $u_{ns}$ and are evaluated only $n_1$ times.
However, for high-dimensional UQ problems, the dimensions for the non-sparse uncertain inputs can still be large enough such that the required model evaluation cost is still unaffordable even with the AMTC method.
We address this problem by applying the AS-NIPC method within the space of $u_{ns}$ to generate an efficient quadrature rule with respect to the active variables in the space non-sparse uncertain inputs. This involves computing the $C$ matrix as
\begin{equation}
    C = \mathbb{E} [(\nabla_{u_{ns}} f)(\nabla_{u_{ns}} f)^T] = \int_{\Gamma} (\nabla_{u_{ns}} f)(\nabla_{u_{ns}} f)^T\rho(u) du\\,
\end{equation}
then the AS method can be used to find the active variables that satisfy
\begin{equation}
    \Tilde{u}_{ns} = W_1^T u_{ns},
\end{equation}
such that $\Tilde{u}_{ns} \in \mathbb{R}^m$ with $m \leq d_1$. 
Following the AS-NIPC method, we can generate the PCE basis functions of the active variables, written as $\Phi(\Tilde{U}_{\{ns\}})$ as well as the quadrature rule with the nodes in $U_{ns}$ space, we denote the nodes as $\boldsymbol{u}_{ns}^{\Tilde{n}}$.
Using the NIPC approach, we can approximate the stochastic output as the PCE basis functions of the active variables and the non-sparse uncertain inputs, as
\begin{equation}
        f(\Tilde{U}_{ns}, U_{s}) \approx \sum_{i = 0}^{q} \alpha_i \Phi_i(\Tilde{U}_{ns}, U_{s}),
\end{equation}
where the PCE basis functions can be constructed as
\begin{equation}
\label{eqn: PCE_total}
    \Phi(\Tilde{U}_{ns}, U_{s}) =  \Phi(\Tilde{U}_{ns}) \times \phi(u_{s_1}) \times \ldots \times \phi(u_{s_{d_2}}),
\end{equation}
where $\phi(u_{s_i})$ represent the univariate orthogonal polynomials of $u_{s_i}$.
The quadrature rule follows the same tensor structure and the quadrature points can be written as
\begin{equation}
\label{eqn: qp_total}
\boldsymbol{u} = 
\boldsymbol{u}^{\Tilde{n}}_{\{ns\}}
\times
\boldsymbol{u}^{k}_{\{s_1\}}\times \ldots,\boldsymbol{u}^{k}_{\{s_{d_1}\}},
\end{equation}
where the $\boldsymbol{u}^{k}_{\{s_i\}}$ is chosen as the $k$ Gauss quadrature points in $u_{s_i}$ dimension. This results in an efficient quadrature rule with a desired tensor structure to use with the AMTC method when it comes to model evaluations.

\subsubsection{AS-AMTC algorithm}
We refer to this AS-based NIPC method as the AS-NIPC method and present a detailed step-by-step algorithm:
\begin{enumerate}
    \item \textbf{Identify the sparse uncertain inputs:} Identify the sparse uncertain inputs that only affect a small amount of the computational cost.
    \item \textbf{Apply AS-NIPC:} Apply AS-NIPC in the non-sparse uncertain inputs space. Generate the PCE basis functions of the active variables as well as the quadrature rule in the non-sparse uncertain inputs space.
    \item \textbf{Form the tensor-structured quadrature rule:} Form the PCE basis functions and the quadrature rule in a tensor structured following \eqref{eqn: PCE_total} and \eqref{eqn: qp_total}, respectively.
    \item \textbf{Model evaluations with AMTC:}
    Perform the model evaluations on the tensor-structured inputs using AMTC.
    \item \textbf{Compute the QoIs:} Compute the PCE coefficients and the desired quantity of interests of the model output.
\end{enumerate}
\section{Numerical Results}
\label{Sec: Numerical Results}
\subsection{7-dimensional analytical piston simulation problem}
The first test problem involves an analytical, nonlinear model that calculates the cycle time of a piston within a cylinder, adapted from \cite{ben2007modeling}. The cycle time $C$ of the piston, measured in seconds, is defined as:
\begin{equation}
C=2 \pi \sqrt{\frac{M}{k+S^{2} \frac{P_{0} V_{0} T_{a}}{T_{0} V^{2}}}},
\end{equation}
where $V$ is given by:
\begin{equation}
V=\frac{S}{2 k}\left(\sqrt{A^{2}+4 k \frac{P_{0} V_{0}}{T_{0}} T_{a}}-A\right)
\end{equation}
and $A$ is calculated as
\begin{equation}
    A=P_{0} S+19.62 M-\frac{k V_{0}}{S}.
\end{equation}
The UQ problem aims to determine the expected value of the piston cycle time considering seven uncertain inputs, detailed in Table \ref{tab:piston}.
\begin{table}
    \caption{Uncertain inputs and distributions for the piston problem}
    \centering
    \begin{tabular}{c c c}
         Uncertain input & Distribution & Description  \\
         \hline
         $M$ & $N(45, 3)$ & Piston weight (kg) \\
         $S$ & $N(0.01, 0.001)$ & Piston surface area ($m^{2}$) \\
         $V_{0}$ & $N(0.010, 0.001)$ & Initial gas volume ($m^{3}$) \\
         $k$ & $N(3000, 200)$ & Spring coefficient ($N/m$) \\
         $P_{0}$ & $N(90000, 5000)$ & Atmospheric pressure ($N/m^{2}$) \\
         $T_{a}$ & $N(290, 20)$ & Ambient temperature (K) \\
         $T_{0}$ & $N(340, 20)$ & Filling gas temperature (K) \\
    \end{tabular}
    \label{tab:piston}
\end{table}

\begin{figure}
\centering
  \includegraphics[width= 7cm]{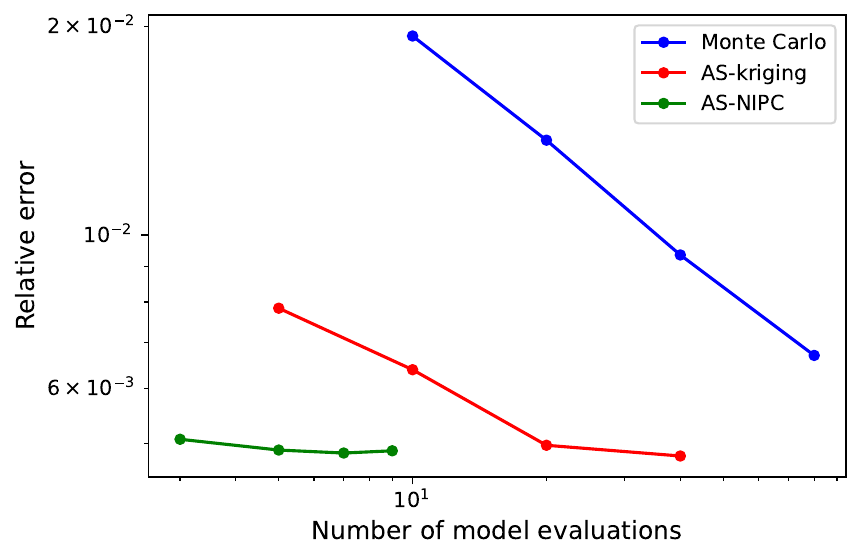}
\caption{UQ results on the piston problem}
\label{fig:piston_results}
\end{figure}

This problem is solved using three UQ methods: Monte Carlo, Active-Subspace-based kriging (AS-kriging), and AS-NIPC. 
The UQ result using Monte Carlo with 100,000 sample points is regarded as the ground truth result.
AS-kriging is implemented by randomly sampling the active variables according to their distributions and constructing the corresponding response surface function. 
To assess their performance, the relative errors of AS-kriging and Monte Carlo methods are averaged over 20 runs.
For AS-NIPC and AS-kriging, the approximation of the $C$ matrix is computed using 100 sample points, and the resulting eigenvalues are presented in Table \ref{tab: piston_eigen}.
\begin{table}[]
\caption{Normalized eigenvalues of the $C$ matrix in the piston problem}
\centering
\begin{tabular}{c} 
 AS-kriging \& AS-NIPC \\ 
 \hline
1.00 \\
3.82e-16   \\
1.53e-21  \\
1.13e-25  \\
1.68e-25 \\
9.71e-27 \\
2.17e-30
\end{tabular}
\label{tab: piston_eigen} 
\end{table}
In this problem, we notice a rapid decline in eigenvalues after the first one, leading us to select a one-dimensional active subspace ($m = 1$). The convergence plots for the three UQ methods are depicted in Figure \ref{fig:piston_results}. 
It is evident from these plots that both AS-based approaches outperform the Monte Carlo method significantly when constrained to 10s or fewer model evaluations. This improvement is attributed to the discovery of a 1D active subspace, effectively reducing the UQ problem's dimension from 7 to 1.
When comparing AS-kriging and AS-NIPC methods, both tend to reach a relative error limit around $5e-3$. 
However, AS-NIPC achieves this level of accuracy with considerably fewer evaluations, owing to NIPC's effectiveness in handling low-dimensional UQ problems. 
Specifically, in this problem, AS-NIPC with only 5 model evaluations results in a lower error than AS-kriging with 20 model evaluations.

\subsection{81-dimensional air-taxi trajectory simulation problem}

\begin{figure}
\centering
  \includegraphics[width= 12cm]{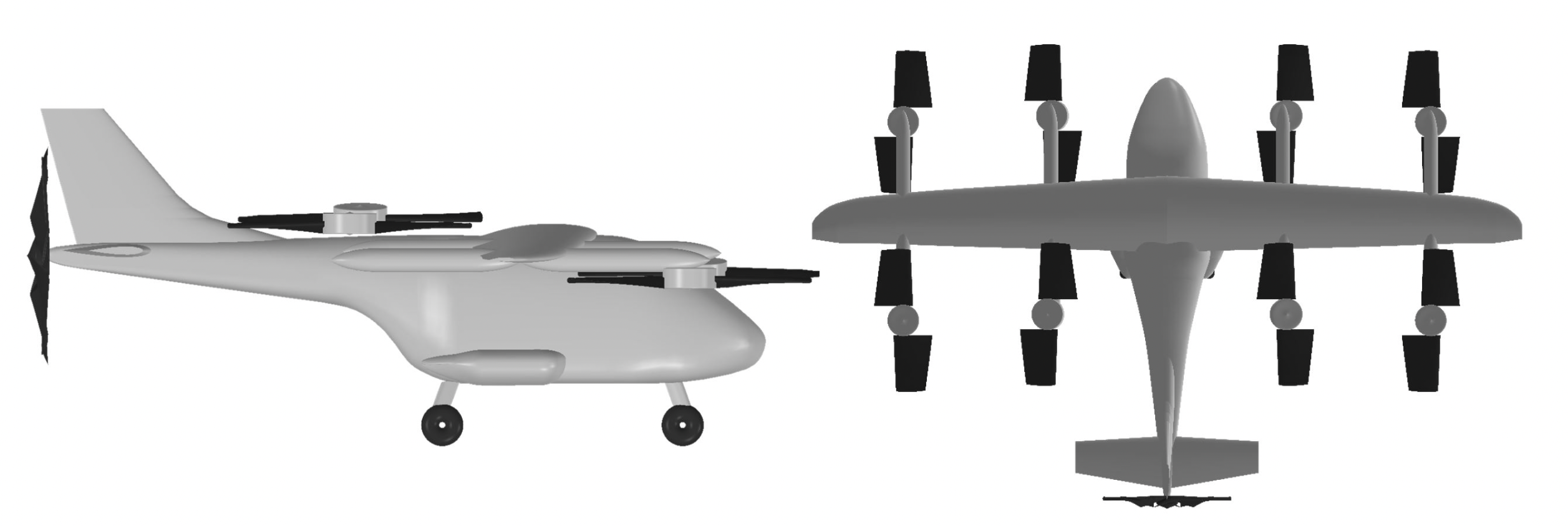}
\caption{Representation of the lift-plus-cruise electric air taxi~\cite{orndorff2023air}}
\label{fig:evtol_graph}
\end{figure}

\begin{figure}
\centering
  \includegraphics[width= 14 cm]{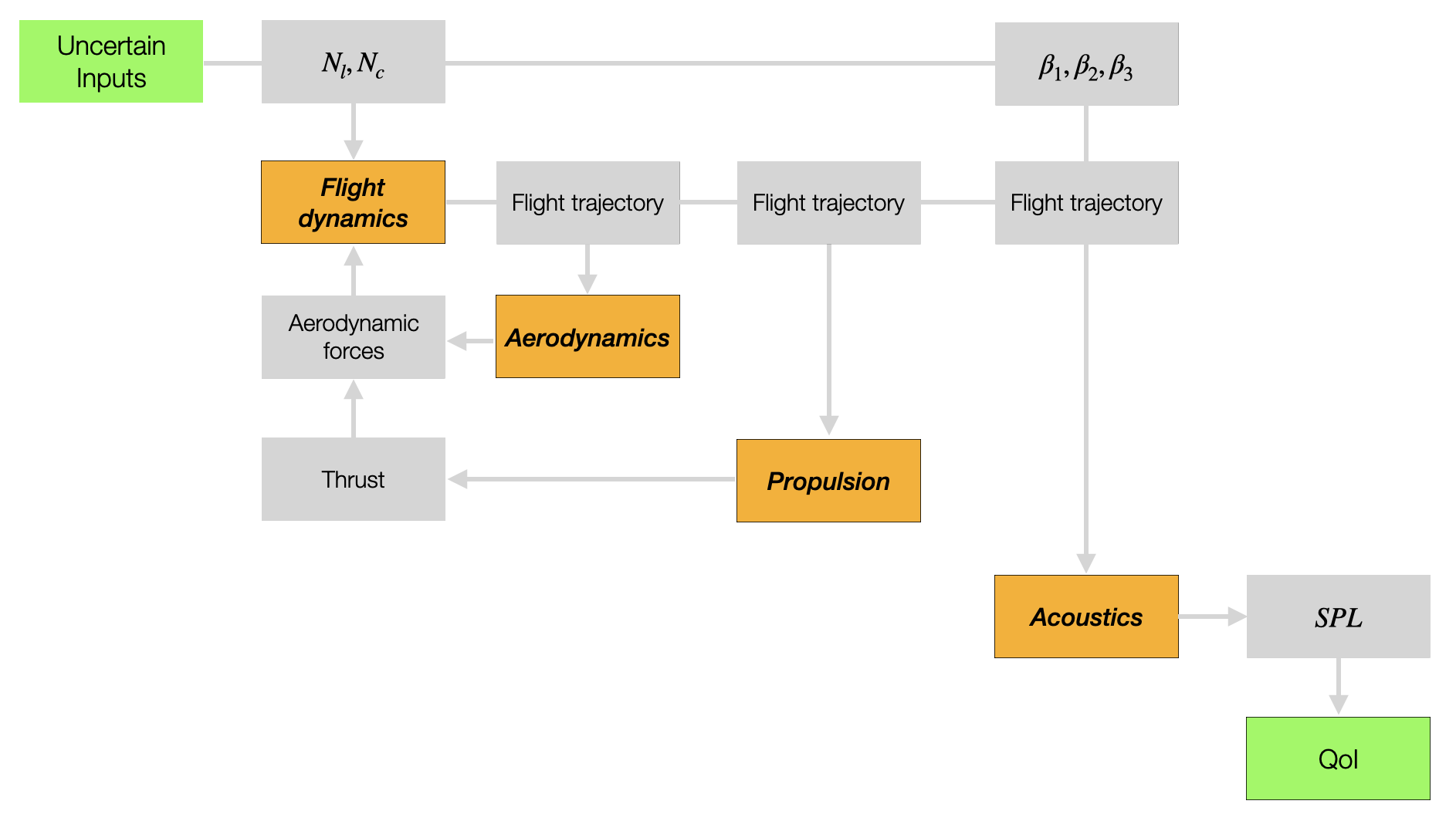}
\caption{Multidisciplinary structure of the air-taxi model}
\label{fig:airtaxi_multi}
\end{figure}

The second test problem is an 81-dimensional UQ problem involving a lift-plus-cruise electric air taxi trajectory simulation model. 
The representation of the aircraft is shown in Fig.~\ref{fig:evtol_graph}. 
This computational model calculates the average ground-level sound pressure level during the aircraft's flight along a specific trajectory~\cite{orndorff2023air}. The UQ problem aims to determine the expected output under uncertainties in the control inputs and acoustic parameters.
The model is composed of two sub-models: the trajectory model and the acoustic model. The trajectory model integrates three disciplines—flight dynamics, aerodynamics, and propulsion—with two-way coupling among them. 
It computes the aircraft's flight path based on the control inputs history, including the histories of lift and cruise rotor thrust ($x_l$, $x_c$).
On the other hand, the acoustics model computes the ground-level sound pressure level (SPL) based on the flight trajectory, using a correlation equation with three parameters: ($\beta_1, \beta_2, \beta_3$). Further details about these models can be found in \cite{orndorff2023air}.
The multidisciplinary structure of the model is shown in ~\ref{fig:airtaxi_multi}.

\begin{figure}%
    \centering
    \subfloat[\centering Cruise rotor thrust]{{\includegraphics[width=7cm]{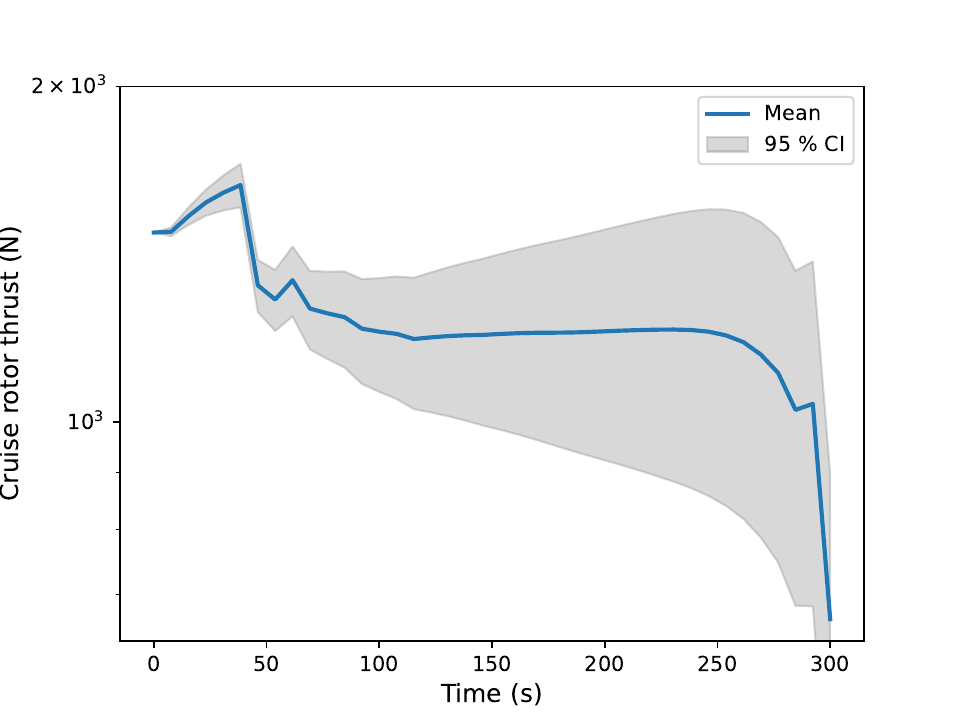} }}%
    \qquad
    \subfloat[\centering Lift rotor thrust]{{\includegraphics[width=7 cm]{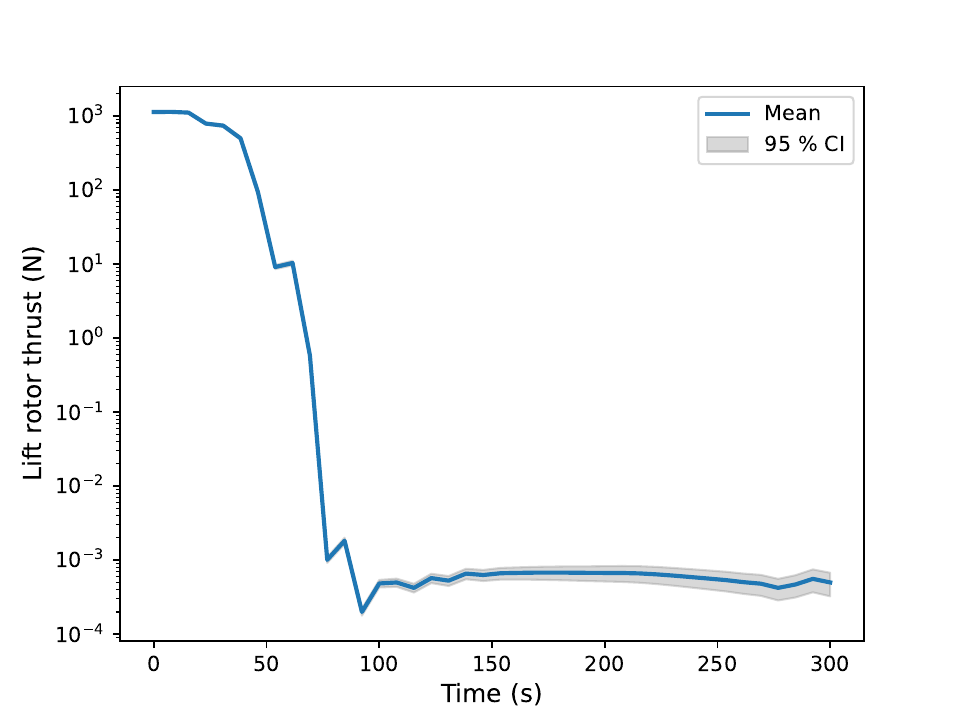}}}%
    \caption{Control inputs with confidence intervals}%
    \label{fig:control inputs}%
\end{figure}

In this UQ problem, we have control inputs with $40$ time steps, denoted as $x_l  =  [x_{l_0}, \ldots,x_{l_{39}}]$ and $x_c  =  [x_{c_0}, \ldots,x_{c_{39}}]$. 
These inputs' average history is denoted as $\Bar{x}_l  =  [\Bar{x}_{l_0}, \ldots,\Bar{x}_{l_{39}}]$ and  $\Bar{x}_c  =  [\Bar{x}_{c_0}, \ldots,\Bar{x}_{c_{39}}]$.
We assume a linear relationship between consecutive steps of control inputs, with white noise added at each step, following:
 \begin{equation}
 \begin{aligned}
     & X_{l_{i+1}} = cX_{l_i} + N_{l_{i+1}}, \quad N_{l_{i+1}} \sim \mathcal{N}(0, 0.003\Bar{x}_{l_i}); \\
     & X_{c_{i+1}} = cX_{c_i} + N_{c_{i+1}}, \quad N_{c_{i+1}} \sim \mathcal{N}(0, 0.003\Bar{x}_{c_i}). \\
 \end{aligned}
 \end{equation}
 This formulation is commonly used to represent control input uncertainties in motion planning under uncertainty problems \cite{janson2017monte}. Fig.~\ref{fig:control inputs} illustrates the control inputs' histories alongside their averages and 95\% confidence intervals.
 Additionally, we have Gaussian uncertain inputs associated with the three parameters in the acoustic model, $(\beta_1, \beta_2, \beta_3)$.
 The complete set of 81 uncertain inputs and their distributions are detailed in Tab.~\ref{tab:uncertain_inputs}.

\begin{table}[h!]
  \begin{center}
    \caption{
    Uncertain inputs in the UQ problem
    }
    \label{tab:uncertain_inputs}
    \begin{tabular}{ c | c } 
    \hline
    Uncertain input & Distribution \\
    \hline
        $N_{l} \in \mathbb{R}^{39}$ &  $\mathcal{N}(0, 0.003\Bar{x}_l)$ \\

        $N_{c} \in \mathbb{R}^{39}$ &  $\mathcal{N}(0, 0.003\Bar{x}_c)$ \\

       $\beta_1\in \mathbb{R} $ & $\mathcal{N}(0.0209, 0.002)$ \\

       $\beta_2\in \mathbb{R} $ & $\mathcal{N}(18.2429, 2)$ \\

       $\beta_3\in \mathbb{R} $ & $\mathcal{N}(6.729, 0.7)$ \\
    \end{tabular}
  \end{center}
\end{table}
This problem is solved by four UQ methods: Monte Carlo, AS-kriging, AS-NIPC, and AS-AMTC. 
Both the AS-kriging and AS-NIPC methods involve applying the AS method in the original uncertain input space. 
For these two methods, the $C$ matrix in \eqref{eqn: 1} is approximated using the Monte Carlo method with the same 100 sample points. 
We present the first seven eigenvalues of the $C$ matrix in Tab.~\ref{tab: eigen} and choose $m = 6$ as the dimension of the active subspace as we observe a rapid decay after the sixth eigenvalue.
For the AS-AMTC method, we first compute the sparsity ratio of each uncertain input and the results are shown in Tab.~\ref{tab: sparsity}. 
The results indicate that the parameter uncertain inputs, ($\beta_1$,$\beta_2$, $\beta_3$) are the sparse uncertain inputs in this computational model with the sparsity ratio of $2\%$. 
This observation can be attributed to the fact that these parameter uncertainties only affect the operations within the acoustic sub-model, which constitutes a very small portion of the overall computational cost.
In AS-NIPC, the AS method is specifically employed within the non-sparse uncertain input space, which comprises only $N_l$ and $N_c$. 
In Tab.~\ref{tab: eigen}, we present the first seven eigenvalues alongside those corresponding to the other two AS methods.
Analyzing Tab.~\ref{tab: eigen}, we notice that the eigenvalues in the AS-AMTC method decay more rapidly compared to the other case, leading us to select an active subspace dimension of $m = 5$.
\begin{table}[]
\caption{Normalized eigenvalues (first seven) of the $C$ matrix}
\centering
\begin{tabular}{c | c} 
 AS-kriging \& AS-NIPC & AS-AMTC \\ 
 \hline
 1.0000 & 1.0000 \\
 0.0661 & 0.0592   \\
 0.0649 & 0.0381  \\
 0.0459 & 0.0080  \\
 0.0346 & 0.0032 \\
 0.0194 & 0.0004 \\
 0.0032 & 0.0002
\end{tabular}
\label{tab: eigen} 
\end{table}

\begin{figure}%
    \centering{{\includegraphics[width=7cm]{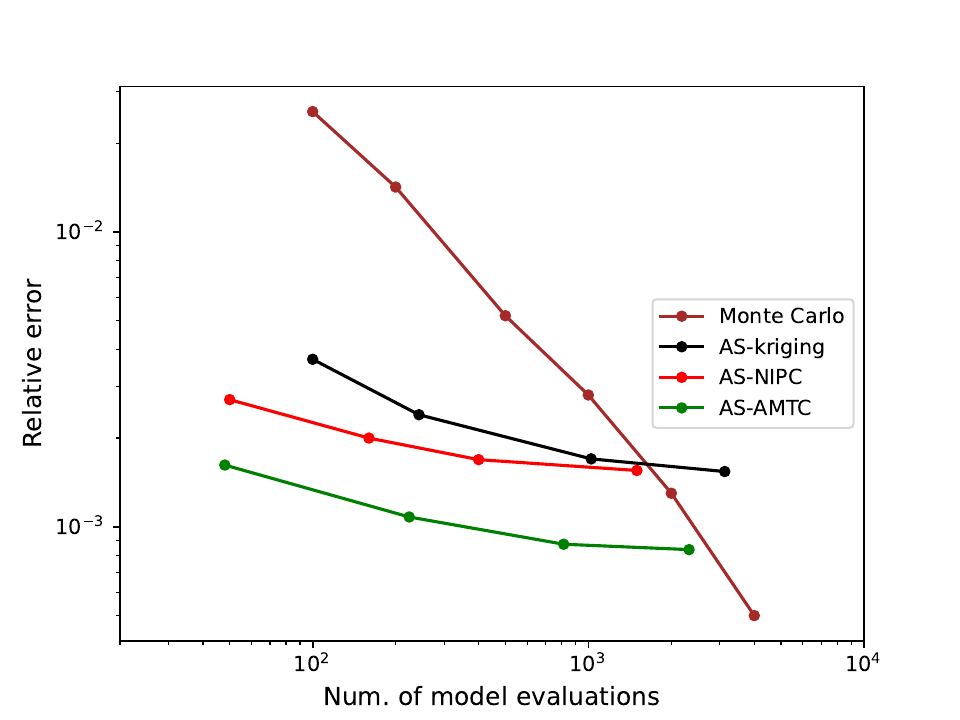}}}%
    \caption{Convergence plot of the four UQ methods}%
    \label{fig:result_81}%
\end{figure}

\begin{table}[]
\caption{Sparsity ratio of the uncertain inputs}
\centering
\begin{tabular}{c | c} 
 Uncertain inputs & Sparsity ratio \\ 
 \hline
$N_{l}$ & 99\% \\
$N_{c}$ & 99\%   \\
$\beta_1$ & 2\%  \\
$\beta_2$ & 2\%  \\
$\beta_3$ & 2\% \\
\end{tabular}
\label{tab: sparsity} 
\end{table}
The convergence plots of the four UQ methods are shown in Fig.~\ref{fig:result_81}. 
Notably, the Monte Carlo method exhibits superior performance when conducting $3000$ or more model evaluations. 
This superiority stems from the inherent trade-off in dimension reduction methods, where reducing problem dimensions results in information loss about the inputs, limiting the dimension reduction-based UQ methods' ability to achieve extremely high accuracy.
When compared to the AS-kriging method, both of our proposed methods demonstrate superior performance. For instance, with 100 function evaluations, AS-NIPC achieves a relative error at least 30\% lower than AS-kriging, while AS-AMTC achieves a relative error at least 80\% lower than AS-kriging. The exceptional performance of AS-AMTC can be attributed to the presence of three sparse uncertain inputs in the computational graph of the model.
These inputs only affect about 2\% of the model evaluation cost.
In this case, maintaining a tensor structure of the quadrature rule with each of the sparse uncertain inputs is very effective to use with the AMTC method. 
Additionally, by applying the AS method exclusively to the uncertain input space excluding the sparse uncertain inputs, we observe a more rapid decay rate of the eigenvalues for the $C$ matrix, as detailed in Tab.~\ref{tab: eigen}. This enables us to choose a smaller dimension for the active subspace while achieving a more accurate reduced-order model.

\section{Conclusion}
\label{Sec: Conclusion}

In this paper, we present two UQ methods, AS-NIPC and AS-AMTC, for solving high-dimensional UQ problems. The AS-NIPC method integrates the integration-based non-intrusive polynomial chaos method with the active subspace method.
This method involves generating the PCE basis functions of the active subspace uncertain inputs and the corresponding quadrature rules to estimate the PCE coefficients.
Expanding upon AS-NIPC, the AS-AMTC method further incorporates the computational graph transformation method, AMTC.
This extension involves generating a desired tensor structure of the quadrature rule and accelerating the tensor-grid model evaluations by using AMTC. 
In a 7-dimensional UQ problem involving a nonlinear piston simulation model, AS-NIPC demonstrates superior efficiency compared to the existing methods.
In an 81-dimensional UQ problem involving a multidisciplinary air-taxi trajectory simulation model, both AS-NIPC and AS-AMTC outperform existing methods. AS-NIPC achieves a 30\% reduction in relative error, while AS-AMTC achieves an 80\% reduction in relative error, exemplifying their exceptional performance in high-dimensional UQ problems.

An inherent limitation of the active subspace method is its ability to identify only those active variables that are linear transformations of the original input variables, restricting its utility in certain scenarios. Thus, future research could explore the dimension reduction techniques capable of discovering active variables that are nonlinear transformations of the original inputs.
Another interesting direction is to explore the performance of the proposed methods in addressing optimization under uncertainty (OUU) problems. A particular challenge lies in developing strategies to manage computational models with multiple outputs in OUU problems, such as objective function outputs and constraint function outputs.

\section{Acknowledgments}

The material presented in this paper is, in part, based upon work supported by  DARPA under grant No.~D23AP00028-00 and by NASA under award No.~80NSSC21M0070

\section*{Declarations}
\subsection*{Conflict of interest} The authors declare that they have no known competing financial interests or personal relationships that could have appeared to influence the work reported in this paper.

\subsection*{Replication of results} The numerical results presented in this document
can be replicated using the methodology and formulations described
herein. The data used in the test problems are available upon request to the
corresponding author.

\bibliography{bibi}
\end{document}